# Superconducting magnet designs and MRI accessibility: a review

Marina Manso Jimeno[1,2], John Thomas Vaughan[1,2], Sairam Geethanath[2,3]*

[1] Department of Biomedical Engineering, Columbia University in the City of New York, 10027, New York, NY, USA

[2] Columbia Magnetic Resonance Research Center, Columbia University in the City of New York, New York, 10027, NY, USA

[3] The Biomedical Engineering and Imaging Institute, Department of Diagnostic, Molecular and Interventional Radiology, Icahn School of Medicine at Mount Sinai, New York, 10029, NY, USA

**Corresponding author:**
Sairam Geethanath
sairam.geethanath@mssm.edu
1470 Madison Ave,
New York, NY, 10029

## Abstract

Presently, Magnetic Resonance Imaging (MRI) magnets must deliver excellent magnetic field ($B_0$) uniformity to achieve optimum image quality. Long magnets can satisfy the homogeneity requirements but require considerable superconducting material. These designs result in large, heavy, costly, and unstable systems that aggravate as field strength increases. This issue is a crucial factor in the disparity of MR density and field strength use across the globe. Low-income settings show reduced access to MRI, especially to high field strengths. This article summarizes the proposed modifications to MRI superconducting magnet design and their impact on accessibility, including compact, reduced liquid Helium and specialty systems. Reducing the amount of superconductor inevitably entails shrinking the magnet size, resulting in higher field inhomogeneity. This work also reviews the state-of-the-art imaging and reconstruction methods to overcome this issue. Finally, we summarize the current and future challenges and opportunities in the design of accessible MRI.

# Abbreviations

**ADC**

Apparent Diffusion Coefficient

**AI**

Artificial Intelligence

**BSCCO**

Bismuth Strontium Calcium Copper Oxide

**DL**

Deep Learning

**DSV**

Diameter of Spherical Volume

**DTI**

Diffusion Tensor Imaging

**DWI**

Diffusion Weighted Imaging

**C3T**

Compact 3T

**EPI**

Echo-Planar Imaging

**FA**

Fractional Anisotropy

**FEM**

Finite Element Method

**fMRI**

Functional Magnetic Resonance Imaging

**FOV**

Field of View

**FSE**

Fast Spin Echo

**GA**

Genetic Algorithm

**GRE**

Gradient Recalled Echo

**HTS**

High Temperature Superconductor

**LP**

Linear Programming

**MAVRIC**

Multi-Acquisition Variable-Resonance Image Combination

**MAVRIC-SL**

Multi-Acquisition Variable-Resonance Image Combination Selective

**MgB$_2$**

Magnesium diboride

**MP-SSFP**

Missing Pulse Steady-State Free Precession

**MRI**

Magnetic Resonance Imaging

**MSI**

Multi-Spectral Imaging

**MQE**

Minimum Quench Energy

**NbTi**

Niobium Titanium

**NZPV**

Normal Zone Propagation Velocity

**PNS**

Peripheral Nerve Stimulation

**ppm**

Parts per Million

**p-p**

peak-to-peak

**REBCO**

Rare Earth Barium Copper Oxide

**REMODEL**

RE- construction of MR images acquired in highly inhOmogeneous fields using DEep Learning

**RMSE**

Root Mean Squared Error

**RF**

Radio Frequency

**SAR**

Specific Absorption Rate

**SEM fields**

Spatial Encoding Magnetic fields

**SEMAC**

Slice Encoding for Metal Artifact Correction

**SNR**

Signal-to-Noise Ratio

**SQUID**

Superconducting Quantum Interference Device

**SWI**

Susceptibility-Weighted Imaging

**TE**

Time of Echo

**Vrms**

Volume root mean square

**WHO**

World Health Organization

**YBCO**

Yttrium Barium Copper Oxide

**ZBO**

Zero Boil-Off

## 1. Magnetic field homogeneity vs. accessibility

Magnetic Resonance Imaging (MRI) is a noninvasive imaging modality that offers good soft-tissue contrast, leveraged for clinical benefits. The goals of improving its spatial resolution and signal sensitivity have driven it toward higher main magnetic field ($B_0$) strengths since its early days in the mid-1970s[1]. The first whole-body superconducting scanner had a field strength of 0.15T[2]. In contrast, conventional clinical field strengths today are 1.5T and 3T[1–3]. In addition to these widely used magnetic field strengths, scanners at 7T[4] and 10.5T[5] are available, while even higher strength scanners are being developed[1,2,5]. The quality of MR images depends on innumerable factors and acquisition parameters but most notably on the spatial uniformity of the main magnetic field[6,7]. However, challenges related to field inhomogeneity are more significant As $B_0$ increases[8]. Accurate spatial encoding mandates tight requirements on field homogeneity to prevent artifacts in the images. Precisely, the field uniformity of a whole-body magnet after installation must be in the order of 10 parts per million (ppm) peak-to-peak (p-p) over a 45-50 cm Diameter of Spherical Volume (DSV)[9–11]. Some imaging techniques even require $B_0$ homogenization methods to further reduce this value below 2 ppm p-p over the Field of View (FOV) during image acquisition[10].

Minimizing field deviations and conductor volume are the primary objectives of magnet design algorithms. The optimization process is iterative and integrates the solution of a Linear Programming (LP)[12,13] or Genetic Algorithm (GA)[14–17] with a Finite Element Method (FEM) simulation software. The algorithm returns the optimum values for parameters such as the magnet's number of coils, the number of turns per coil, their spatial coordinates, and dimensions. Subsequently, the simulation software inputs these parameters to guarantee that the resultant magnetic field distribution fulfills the homogeneity requirements. Theoretically, only an infinitely long solenoid can achieve a perfectly homogeneous field. Due to the unfeasibility of such configuration, algorithms set constraints on final magnet dimensions, weight, and cost depending on the target application[12]. However, the trade-off between field uniformity and magnet length often results in exceedingly large and heavy MRI scanners. The amount of superconductor required to manufacture such magnets impacts the final product's cost and hinders the systems' transportation and siting options. In addition, other MRI components such as the gradient, shimming and Radio Frequency (RF) coils, and refrigeration system further increase the scanner's complexity, power requirements, dimensions, and cost[18].

Figure 1 illustrates the whole-body superconducting magnet manufacturing trend over the past three decades. On the one hand, the push toward higher field strengths is perceptible. During the 1990s, 0.5T and 1T scanners were prevalent, and the first 1.5T scanners emerged. Around the 2000s, the first 3T arose in conjunction with compact 1.5T systems, replacing lower field strength scanners as the clinical standard[1–3]. However, the magnet's required amount of superconductor increases with field strength. Consequently, the scanner complexity and previously mentioned cost and siting challenges also grow. For instance[1], the length and weight of a 60 cm bore 1.5T clinical MRI are 1.71 m and 4.5 tons, respectively. It requires a minimum space of 28 $m^2$ and less than 1500 L of liquid Helium for cooling purposes[19]. In contrast, the first available clinical 7T scanner is 2.97 m long, weighs approximately 20 tons, demanding a 65 $m^2$ room and 4000 L of cryogen[20]. On the other hand, Figure 1 portrays how manufacturers have progressively reduced the magnet coil volume relative to early models thanks to innovations in optimization algorithms and superconducting materials. Compact and lightweight designs achieve two fundamental goals: reducing the scanner's cost and improving patient comfort. Additionally, they improve MR accessibility, which has become a critical objective for superconducting and permanent magnet design.

The price of an MRI system is roughly $1M per Tesla (T)[21], to which installation, operation, and maintenance costs add. This imaging technique provides remarkable image quality and continues to unfold new diagnostic possibilities. However, most scanners do not comply with the accessibility dimensions introduced by Geethanath & Vaughan[22]. Their elevated cost and exigent infrastructure requisites have led to a heterogeneous distribution of MRI technology across the globe[22–24]. While reports on the disparity of MRI density across world regions are available[25], it is crucial to further characterize this discrepancy according to field strength (Figure 2). Field strengths of 1.5T and above represent 85% of the MRI market in the US and Europe[26]. Conversely, scanners below 1.5T are still the most abundant in low-resource settings[23,27]. This article considers high and low resource settings based on the World Health Organization (WHO) income level classification[28]. The data illustrated in Figure 1, while limited in availability, represents the current state of scanner density. In this instance, low and lower-middle-income countries exhibit low MR densities and frequently coincide with higher proportions of low field strength systems. These units typically correspond to permanent-magnet-based scanners that involve lower acquisition and maintenance costs[9].

MRI magnets must drastically shrink their size to impact MR access. Two proposed methods have been lowering the field strength (i.e., bottom-right magnet design in Figure 1) and loosening the constraints on field homogeneity. While both are viable approaches, they involve reconsidering certain aspects of the imaging process to overcome challenges such as loss of Signal-to-Noise Ratio (SNR) and image distortions. Novel ultra-low, portable permanent magnets[29–32] have adopted both methods. They deliver substantial benefits in portability and cost but entail prolonged scanning times and reduced image quality[33]. Moreover, advanced and critical imaging techniques such as Diffusion-Weighted Imaging (DWI), Susceptibility-Weighted Imaging (SWI), functional MRI, and angiography are not readily feasible at such low field strengths. For these reasons, this work only considers scanners using superconducting magnets between 0.5-3T. Such systems present the most cost-effective configurations capable of delivering optimal image quality, signal sensitivity, and a wide range of contrasts and image applications[9,11].

---

[1]The 1.5T scanner was chosen from the same manufacturer and with the same patient bore size as the 7T system for a fair comparison.

This work reviews the past and current efforts in superconducting magnet design to improve accessibility. For magnets with higher tolerances for field homogeneity, we outline the results of novel imaging techniques that avoid, correct, or mitigate the artifacts that non-uniformities induce in the images. Finally, the last section summarizes our findings and discusses the challenges and opportunities of accessible MRI magnet design.

## 2.   Accessible magnet configurations

The routine clinical whole-body MRI comprises a cylindrical multi-coil, liquid helium bath-cooled Niobium-titanium (NbTi) superconducting magnet. This configuration is the most cost-effective while ensuring optimal B0 homogeneity conditions. Nevertheless, it also raises challenges such as strong dependence on liquid Helium, narrow mechanical tolerances, and limited siting flexibility. Jointly, the magnet and the refrigeration system account for approximately 38% of the scanner's total cost[18]. Hence, tackling these subsystems can vastly impact MR accessibility.

### 2.1. Compact MRI

Patient space and comfort have been the motivation for new magnet configurations. Com- pact systems feature short and wide patient bore while maintaining the system's overall height and weight low[9]. A wide bore leaves extra room within the scanner walls and the patient, while a short bore length permits the head to lay outside the cylinder in most procedures[11,34]. However, to achieve the same field strength, these configurations require more superconductor material and experience higher peak fields than the longer and narrower bore versions[11]. Parizh et al.[35] demonstrated that in order to achieve 10 ppm homogeneity over the DSV and maintain the same stray field, the field strength has to rise by 0.1T per cm cut off of magnet length. Since magnet cost hinges on the superconducting wire type and amount, the magnet designer must carefully assess trade-offs on field strength, field homogeneity, and fringe field.

Bore length ranges from approximately 1.25-1.95 m for 1.5T systems and 1.65-2.13 m for 3T systems. Figure 3 exemplifies the effects of magnet length, bore diameter, and $B_0$ on field homogeneity. Shorter magnets entail increased field inhomogeneity, while long magnets are needed to attain satisfactory field uniformity for higher $B_0$. Ultra-short magnets may even require more than the standard eight coils (6 main coils, 2 shield coils) to fulfill uniformity standards[9,35]. As an example, the 1.5 m long MAGNETOM Avanto (2003)[36] needed 7 main coils to achieve 0.2 Volume root-mean-square error (Vrms) ppm homogeneity at 40 cm DSV, resulting in a higher weight than similar systems (Figure 3). The shortest bore length observed in a commercial scanner was 1.25 m for the 1.5T MAGNETOM Espree (2004)[37]. The advantages of this scanner are higher patient acceptance and success rates for claustrophobic patients[38,39] and interventional procedures[40]. Nevertheless, these advantages come at the expense of smaller maximum FOV (45x45x30cm$^3$), increased probability of geometric distortion, and longer acquisition times compared to long bore systems of the same field strength[9,39,41].

The quest for openness moved the standard patient bore from a 60 cm to a 70 cm diameter. The advent of the so-called "wide-bore" MRI in 2004, and most new MRI installations present this configuration[35]. The widest patient bore available corresponds to the MAGNETOM Free.Max (2021)[42], with a diameter of 80 cm. However, this broad bore diameter constraint inflicted a trade-off in the system's field strength and gradient specifications. To restrain the overall cost and power consumption, the manufacturers had to reduce the field strength to 0.55T and limit the gradient's maximum slew rate to 127 T/m/s[43]. Wide-bore magnets tend to be also short to limit the amount of superconducting material, compromising $B_0$ homogeneity. Xu et al.[44] determined that decreasing the patient bore diameter is the best method to reduce magnet cost and minimum magnet length. The MAGNETOM Free.Star[45] has adopted this approach. It utilizes the same magnet as the Free.Max but has a 60 cm patient bore, which allows cost reductions and increases accessibility (510k approval pending).

## 2.2. Reduced liquid Helium MRI

NbTi is a mature, mechanically robust, manufacturing-friendly superconductor material optimized for MRI production. However, its low critical temperature of 9.3 K requires operation at liquid Helium temperature, which results in a higher refrigeration and installation cost[35]. Liquid Helium is a rare gas extensively used in industry and paramount in the field of superconducting magnets. Its high demand and the limited number of suppliers have led to increased and fluctuating prices and uncertainty about its future availability[18]. Furthermore, the lack of liquid Helium sources is a crucial cause of the reduced access to high field MRI in remote areas and developing countries[46,47]. The regular NbTi liquid Helium-bathed magnets operate at 4.2 K and contain approximately 1500-2000 liters of liquid Helium[48]. Early systems required periodical cryogen refills, which progressively became more spaced as cryocooler innovations reduced their consumption per hour. Presently, MRI utilizes the "Zero Boil-Off" (ZBO) cooling technology. The cryocooler recondenses the Helium gas that evaporates from the cryogen bath, creating a closed-loop system and eliminating the need for any refill except after a quench event[9,18]. Despite eliminating the cost and burden of the refill operation, the magnet still requires 500 or more liters of liquid Helium, and the cold head has to be periodically monitored and replaced. Specifically, the average cold head lifespan is 4-5 years for a new system and 3-4 years for a refurbished system[49].

Recently, the advances of Gifford-McMahon two-stage cryocoolers have allowed conduction cooling of superconducting magnets. These so-called 'dry' magnets eliminate the liquid Helium bath. Instead, the cold head performs magnet refrigeration via thermal conduction of typically copper straps. Additionally, conduction-cooled magnets are fully sealed, and they do not require the construction of a venting pipe. This attribute allows for a considerably more flexible and affordable siting of these systems. Commercially available examples of this magnet configuration are Philips's BlueSeal 1.5T (2018)[50] and Siemens' DryCool 0.55T (2021)[51]. Both NbTi-based systems reduced their liquid Helium use to 7 and 0.7 L, respectively. However, the low Minimum Quench Energy (MQE) and tight temperature margin of NbTi (Table 1) make these magnets less stable and require a reduction in operating current. Consequently, conduction-cooled NbTi magnets demand more superconducting material to maintain the same field strength[46,52].

Conduction cooling is more appropriate for cooling superconducting magnets with larger temperature margins, such as those based on High Temperature Superconductors (HTS). The use of HTS for MRI application is a niche area of research in the pursuit of liquid Helium-free accessible magnets. These superconducting materials include $MgB_2$, YBCO, ReBCO, and BSCCO. They all have a higher critical temperature than NbTi and allow higher operating temperatures, eliminating the need for liquid Helium as a cryogen. Furthermore, HTS's high MQE of up to some Joules (J) renders very stable magnets and practically eradicates accidental quenches. However, their Normal Zone Propagation Velocity (NZPV) is several orders of magnitude slower than for NbTi. This parameter measures how quickly the magnet can spread its stored energy during a quench event. If the magnet is not sufficiently fast, hot spots are more prone to occur before the traditional quench protection methods can detect them[46].

The MROpen EVO[53] is currently the only HTS-based commercially available scanner. This $MgB_2$ 0.5T magnet has an open upright configuration and operates in driven mode.

Besides this scanner, in vitro imaging has been demonstrated for a 1.5T YBCO magnet for extremity imaging[54], a 1.5T BSCCO magnet for head imaging[55], and a 3.0T BSCCO small magnet[56]. These prototype magnets were tape wounded and operating at 20 K. There are still pending challenges before manufacturing a viable commercial HTS scanner. These include the conductor properties and price[9,35], availability of joints for persistent operation[57–59], and safe active quench protection systems[46,60].

Table 1 summarizes the most relevant characteristics of potentially accessible systems, comparing them among the standard liquid Helium-bathed NbTi magnet, new conduction-cooled NbTi magnets, and prototype conduction-cooled HTS-based magnets. Most of the properties for the latter configuration correspond to $MgB_2$-based-magnets, as Parizh et al.[35] concluded that this HTS has the best success probabilities in their extensive review of superconductors beyond NbTi.

## 2.3. Specialty MRI

When compactness is a priority but field homogeneity must be preserved, system designers may reduce the imaging volume and limit the system's application to dedicated examinations[9]. These specialty MRI systems are anatomy-targeted to different body regions such as the head, extremities, breast, or imaging of neonates. Accordingly, the design of the system undergoes modifications tailored to its application.

The Compact 3T (C3T) head-only system initially introduced by Foo et al.[61] is another example of a sealed conduction-cooled NbTi magnet that requires only 12 L of liquid Helium. This system demonstrated safe brain imaging using an optimized gradient system with 80 /m amplitude and 700 T/m/s slew rate[62]. The slew rate increase allowed for notably shorter TE in DWI and Fast Spin Echo (FSE) and echo spacing reduction in Echo-Planar Imaging (EPI) sequences. These reductions in acquisition achieved images with better SNR, sharpness, and geometric fidelity than images acquired in a whole-body 3T scanner[62,63]. Nonetheless, C3T images attained lower Temporal SNR in resting-state fMRI acquisitions, required higher-order gradient nonlinearity correction, and experienced similar amounts of motion artifacts.

The Synaptive 0.5T MRI[64] is another conduction-cooled head-only alternative[65]. Although at a lower $B_0$, its gradient system offers upgraded performance with maximum gradient amplitude and slew rate of 100 /m and 400 T/m/s[66], respectively. DWI and Diffusion Tractography Imaging (DTI) are feasible in this system with similar image quality, Fractional Anisotropy (FA), and Apparent Diffusion Coefficient (ADC) mean values compared to whole-body 1.5T scanners[67–69]. Additionally, the lower field strength allows a 9 and 36-fold drop in Specific Absorption Rate (SAR) compared to 1.5T and 3T[66], respectively.

Various dedicated 1.5T breast MRI scanners are commercially available (Aurora[70], EMMA 1.5T[71]). These units include more powerful gradients and breast-tailored RF coils similar to head scanners. Research using Aurora reported better performance in breast cancer screening than whole-body 1.5T scanners[72]. However, breast-dedicated MRI entails large magnets as they must accommodate the entire torso inside the bore. Both models[70,73] weigh approximately 3 tons and need a minimum room of 55 m$^2$, presenting similar siting challenges as 1.5T whole-body scanners.

GE launched its 1.5T Optima MR430s orthopedics-dedicated scanner in 2011. The magnet has a 21.8 patient bore, weighs approximately 400 kg, and requires only 50 L of liquid Helium. The reduced gradient size provided low power deposition and reduced noise during acquisition. It also practically eradicates Peripheral Never Stimulation (PNS) probability while still delivering 70 /m amplitude and 300 T/m/s slew rate[74]. These characteristics made the system suitable for converting into a neonatal scanner in a clinical setting[75].

In most cases, the efficient use of the bore space renders more compact magnets with reduced costs and footprints and more powerful gradients. These size characteristics facilitate effortless siting, operation, and maintenance, while the gradient systems boost image quality. The reduced size also implies less liquid Helium required for operation, making specialty MRI apter for conduction cooling and HTS magnets[11]. However, their implementation in traditional clinical settings is scarce due to their lack of clinical universality compared to whole-body scanners[35].

## 3. Imaging in highly inhomogeneous fields

Differences in magnetic susceptibility at tissue interfaces or phase accumulation during long readouts can cause field deviations. The effects of these field non-uniformities on the images are well-known; they are responsible for geometric distortions in EPI, blurring in spiral imaging, and signal loss or pile-up. These inhomogeneities are local and relatively small in magnitude, and post-processing correction techniques can tackle the artifacts that they produce. These mitigation methods[76–80] are widely available in open-source packages[81–83] for fMRI and DWI data processing pipelines[84–86]. In contrast, the field inhomogeneity that an accessible, compact magnet pose may involve smoothly varying bandwidths of tens of kilohertz over the imaging volume. The employment of inhomogeneous fields is a contrasting change in the conventional imaging methods but is required to augment MRI accessibility[18]. The challenges that such B$_0$ inhomogeneity poses to the imaging process and the available sequences to overcome them were recently reviewed by Mullen & Garwood[7]. Multi-Spectral imaging (MSI) methods have shown promising potential among the reviewed sequences. These techniques assume large and rapidly varying field inhomogeneities as their purpose is imaging near metallic implants. Some of these sequences are Multi-Acquisition Variable-Resonance Image Combination (MAVRIC)[87], Slice Encoding for Metal Artifact Correction (SEMAC)[88], or the combination of both (MAVRIC-SL)[89].

Severe $B_0$ inhomogeneity poses issues to spatial selection and encoding. These include $T_2$* local dephasing in gradient-echo-based sequences and geometric distortions in the frequency encoding direction and slice profile. Reducing the echo time as much as possible is the only mitigation option for the former artifact. In contrast, solutions to the latter require either lowering $B_0$ or increasing the amplitude of the readout gradients and using non-spatially-selective pulses when possible[7]. For instance, pre-polarized ULF systems excite the spins in higher field strength and immediately ramp down the magnet to a lower strength for spatial encoding[90]. Some of these systems utilize Superconducting Quantum Interference Device (SQUID)[91,92] signal detection to circumvent the lower SNR typical of low $B_0$. ULF systems can also feature Spatial Encoding Magnetic fields (SEM)[16,93] in conjunction with pre-polarized fields. This technique leverages linear inhomogeneities in the field to spatially encode the signal. Despite these methods' ground-breaking characteristics and promising potential, they must be adapted to higher field strengths to be of clinical value.

After manufacture, the field homogeneity of a whole-body superconducting magnet is several hundred ppm[6,9,10]. Manufacturers leverage passive and active shimming to reduce this value to clinical homogeneity standards. Passive shimming utilizes the induced magnetic field of diamagnetic and paramagnetic materials. These strategically placed materials correct hardware imperfections but considerably increase the system's weight[94]. In contrast, active shimming uses superconducting coils and reduces the homogeneity to levels acceptable for imaging before data acquisition. Most $B_0$ imperfections within the magnet bore, including subject-specific susceptibility-based distortions, are mitigated by tailoring the coils' current[94,95]. However, bore space for high-order shimming coils is limited within the bore of a compact magnet[16], and they increase superconducting usage and the complexity of the cryostat design. Furthermore, active $B_0$ shimming relies on the field distribution measurement, and inaccuracies in this acquisition can degrade the shim performance[95].

Ideally, a distortion-free image could be reconstructed from a distorted one if the true field map distribution were known[6,95]. However, the signal loss during image acquisition would be non-recoverable. A $B_0$ map is generally acquired using a double-echo Gradient Recalled

Echo (GRE) sequence with short TE. This acquisition and subsequent phase image calculation are subject to errors caused by phase wrapping, eddy currents, low SNR, and coil-combination methods, even in a homogeneous field. It also involves a longer scanning time due to the extra sequence acquisition during which motion can occur and cause registration errors. Furthermore, when acquired in an inhomogeneous magnetic field, geometric distortion along the frequency encoding direction and spin dephasing typical of GRE sequences may exacerbate the problem, limiting the field map's accuracy and the subsequent artifact correction of the images. Thus, techniques that estimate the field's spatial distribution using alternative methods are essential.

Dual polarity encoding techniques first introduced by Chang & Fitzpatrick[80] estimate the field map from two sets of images acquired with opposite phase encoding directions. Mullen & Garwood[96] used dual polarity in conjunction with a Missing Pulse Steady-State Free Precession (MP-SSFP) sequence to obtain and correct images of a phantom with a metallic implant that mimics field inhomogeneities. They compared the results against images of a MAVRIC sequence, which also provides a $B_0$ map estimate[89,97]. The dual polarity MP-SSFP sequence used low SAR, 20 kHz RF-pulse bandwidths for excitation and refocusing and achieved an acquisition time 3.17-fold faster than single encoded polarity MAVRIC. They reconstructed artifact-free images after correction using the estimated field map, which showed high fidelity to theoretical simulations. The $B_0$ maps indicated off-resonance levels of ±47 ppm (3 kHz at 1.5T). A similar method[98] utilized an unsupervised neural network to obtain the frequency field maps from 3D-MSI dual polarity images. The authors leveraged the predicted field maps to correct pile-up and ripple artifacts near metallic implants that caused off-resonance of up to ±156 ppm (10 kHz at 1.5T).

A patented report[99] of $B_0$ map estimation using an iterative algorithm is also available. The invention's theoretical basis relies on the decomposition of a $B_0$ map when a sample is present in the magnet's theoretical field distribution and the sample-specific perturbations. The algorithm aims to estimate this true field map starting from the artifact-corrupted images and the empty magnet $B_0$ distribution. This field map is constant and requires a single acquisition. In the first iteration, the correction of the distorted image uses the system's field map. For the rest of the iterations, the algorithm updates the field map based on the previous iteration's correction performance. Eventually, the algorithm stops when the corrected image meets some criteria. The inventor claims this method can compensate artifacts on images acquired on magnetic fields with non-uniformities of up to 160 ppm.

A deep-learning-based approach by Gowda et al.[100] introduced REMODEL (RE-

construction of MR images acquired in highly inhOmogeneous fields using DEep Learning). This CNN-based tool inputs corrupted complex k-space data and outputs artifact-free image magnitude data. They tested the model on simulated data which included $T_1$-weighted images corrupted with random-generated field maps of up to ±390 ppm (50 kHz at 3.0T). REMODEL demonstrated faithful reconstruction and Root Mean Squared Error (RMSE) smaller than 0.15 with respect to the ground truth images in all cases.

Figure 4 provides a visual summary of the timeline of methods for imaging in inhomogeneous fields since 1988. It comprises acquisition, $B_0$ map estimation, and reconstruction approaches.

## 4. Challenges and opportunities

Accessible MRI requires significant changes to the whole scanner and imaging process as we currently know them[7,18,48]. As mentioned in Section 1, superconducting cylindrical magnets are the most cost-effective viable option for an accessible MR. In particular, mid-field (0.5-1.5T) scanners can deliver sufficient image quality for clinical use while limiting accessibility challenges. Field strengths higher than 1.5T offer SNR and image resolution advantages but entail increased cost, weight, size, and cooling requirements[9].

The past attempts to modify MRI design have primarily focused on enhancing image quality or patient comfort. Open systems accomplish the latter by better accommodating claustrophobic and obese patients at the cost of more challenges during siting and limitations on field strength[101–103]. In contrast, anatomically-targeted scanners can deliver high performance with a reduced footprint and cost when image quality is the priority. However, due to their lack of clinical universality[18,35], the purchase of several dedicated systems would be required to perform the most common imaging procedures, rendering an impractical approach. The democratization of MRI demands more compact, lightweight, stable scanners that are easy to site and capable of delivering clinical quality performance with low power consumption. Reducing or eliminating the amount of liquid Helium required for MRI operation can ease the siting and stability issues and MRI's dependency on this cryogen[35]. Due to their operating temperature and stability advantages (Table 1), innovations in conduction-cooled HTS magnets will be crucial. However, the feasibility of whole-body HTS still depends on solving some engineering challenges, such as the viability of permanent joints and quench protection systems.

In addition, significantly shrinking the magnet can achieve some of the accessibility requirements. Section 2.1. demonstrated that a wide bore increases the overall scanner's cost and puts pressure on the gradients. Thus, an accessible MR will benefit from a bore narrower than 70 cm. Regarding magnet length, a shorter bore will share advantages with specialty magnets (Section 2.3.). They would be able to house moderately powerful gradients, resulting in improved image quality with reduced PNS. However, it is evident from Section 2.1. that reducing the magnet length leads to an inevitable increase in field inhomogeneity. While $B_0$ shimming methods have been the gold standard to correct these field deviations, these techniques introduce other penalties in the magnet design. Shimming hardware ultimately affects the final weight, bore dimensions, amount of superconductor, and refrigeration requirements (Section 3). Instead, opportunities will unfold if magnets relax their rigid restrictions on field homogeneity[7,18]. Section 3 covered promising methods to acquire artifact-free or artifact-mitigated images in the presence of field inhomogeneities and to estimate the spatial field distribution. Field map synthesis and subsequent image correction using the resulting estimation have been demonstrated feasible analytically and leveraging DL models.

Manufacturers have explored reducing the field strength of superconducting magnets below conventional values to reduce the cost while maintaining field homogeneity constraints. This approach aims to achieve sufficient image quality for clinical use while being conservative on $B_0$. The field strength choice depends on a trade-off between cost, performance, and feasibility of advanced applications. A novel 0.55T scanner has demonstrated the feasibility of this strategy by leveraging the past progress in MRI engineering and Artificial Intelligence (AI)[43]. Additionally, AI has demonstrated outstanding performance in essential tasks such as noise reduction[104–107], artifact detection[108–112] and mitigation[111,113–115], image quality improvement[116–118], and scan automation[119–121]. These capabilities in an accessible MRI system will allow shorter scan times without compromising clinical quality and reduce the requirements for on-site skilled personnel.

In conclusion, an accessible MRI requires a breakthrough in magnet design and imaging techniques. Based on the accessibility demands and a review of the characteristics of existing magnets, we consider that the best candidate will be a whole-body superconducting, mid-field, conduction-cooled HTS-based, short magnet. Such a scanner will also leverage AI methods to optimize image quality, signal sensitivity, ease of operation, and patient throughput.

Table 1. Characteristics of three different magnet configurations using Low versus High Temperature Superconductors (LTS and HTS) and conventional Liquid Helium bath versus conduction cooling assuming a field strength of 1.5T.

| Characteristic | (A) Liquid Helium-bathed NbTi magnets | (B) Conduction-cooled NbTi magnet | (C) Conduction-cooled HTS magnet |
|---|---|---|---|
| **Refrigeration** | | | |
| Liquid Helium capacity (L) | ~1500 | ~7 | ~1 |
| Liquid Helium refills | In the event of quench | NA | NA |
| Operating temperature $T_{op}$ (K) | 4.2 | Higher than (A) | 4-20 |
| **Stability** | | | |
| Temperature margin (K) | 1 | 1 | Depends on $T_{op}$ but higher than NbTi |
| MQE (mJ) | 1-10 | 1-10 | Up to several Joules |
| **Quench protection** | | | |
| Protection system type | Passive | Passive | Active (research stage) |
| Quench pipe | Required | NA (sealed magnet) | NA (sealed magnet) |
| **Persistence** | | | |
| Joint resistance ($\Omega$) | $10^{-12}$ | $10^{-12}$ | Must allow persistent operation (research stage) |
| $I_{op}/I_c$ | ~70% | Lower than (A) due to reduced $I_{op}$ | Depends on the material N-value and $I_c$ but usually lower than (A) |
| **Commercialization** | | | |
| Conductor cost | $ | $ | $$$ (expected to decrease if mass produced) |
| kAmp-km | 15-20 | Larger than (A) due to reduced $I_{op}$ | Larger than (A) due to reduced $J_c$ |

**Figures**

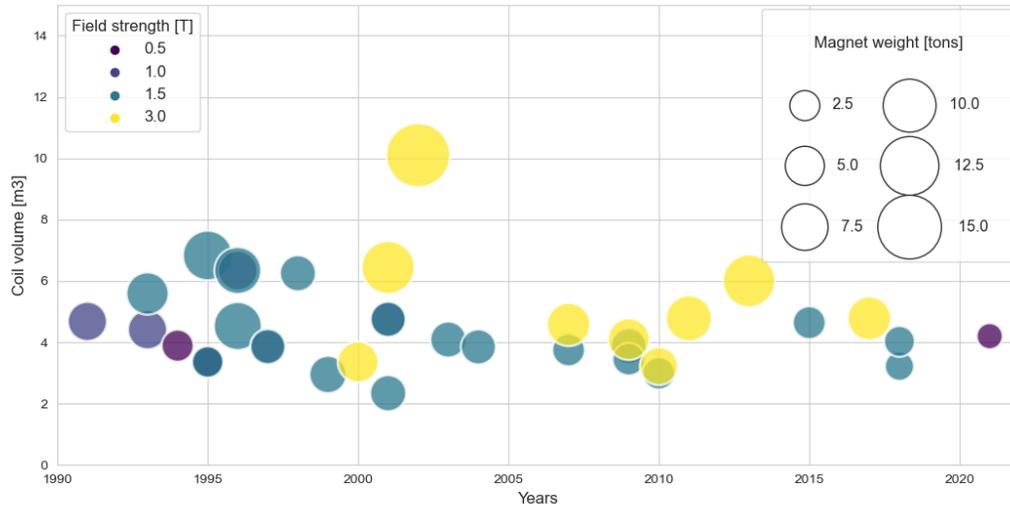

Figure 1. Evolution of whole-body superconducting cylindrical MR magnets coil volume (m³) across the last three decades (years). The magnets are classified by field strength from 0.5-3T and the diameter of each scatter point indicates the magnet's weight (tons). The coil volume is calculated given the patient and warm bore diameter and magnet width and length.

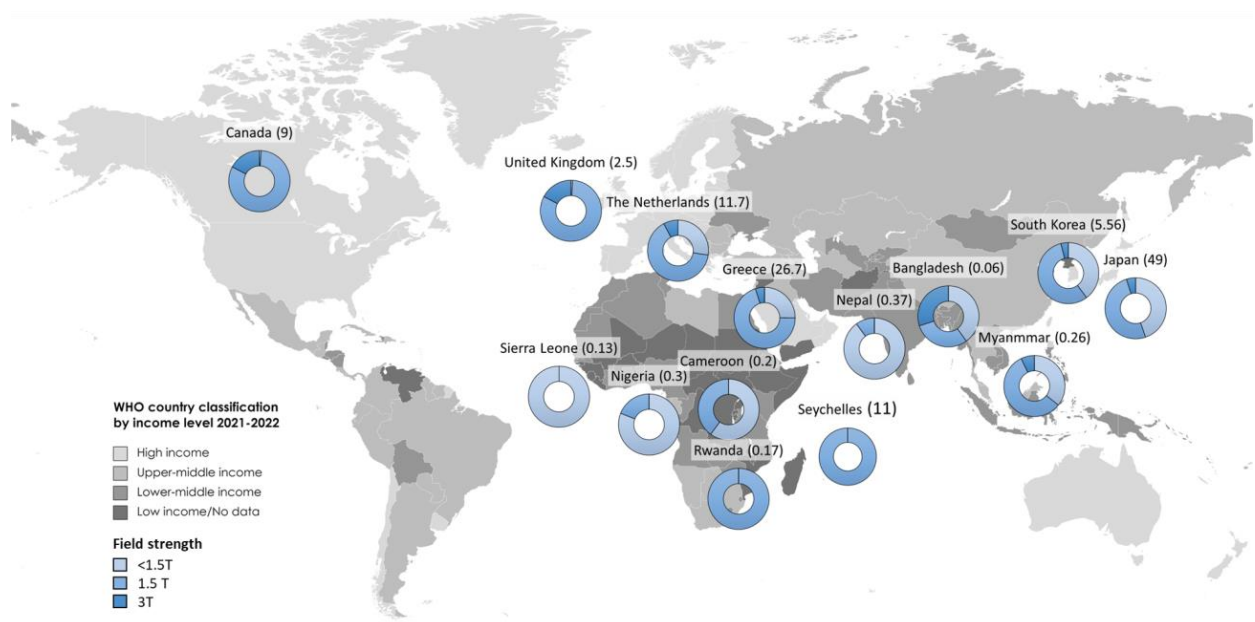

Figure 2. Proportion of MR units classified by field strength in various countries worldwide. The different tones of blue in each wheel chart indicate the proportion of each field strength reported for each country. We considered three categories: below 1.5T, 1.5T, and above 1.5T. Country grayscale coding is based on the World Health Organization's country classification by income 2021-2022. The number in parentheses next to the name of each country is the MR density, i.e., the number of MR units per million inhabitants.

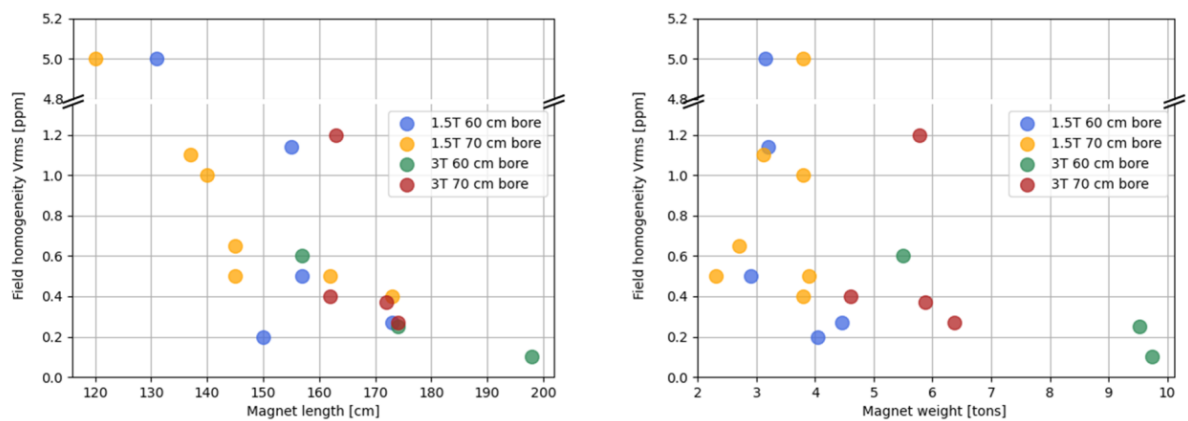

Figure 3. Magnetic field homogeneity (Vrms) at 40 cm Diameter of Spherical Volume (DSV) versus magnet length (left) and weight (right) for different commercially available magnets. The data represents specifications of the magnets used by the most common MRI vendors, classified according to patient bore diameter (60 and 70 cm) and field strength (1.5T and 3T).

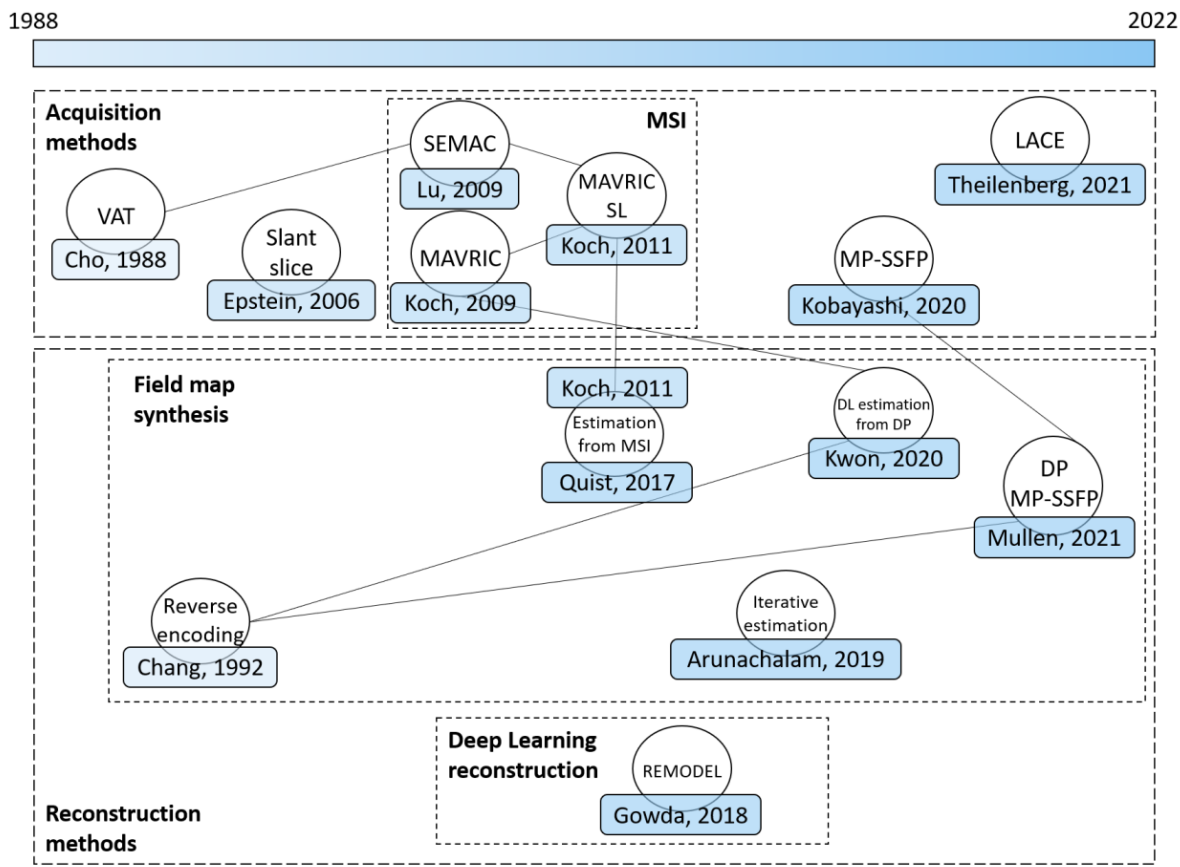

Figure 4. Review of image acquisition and reconstruction in the presence of inhomogeneous fields from 1988 to 2022. Reconstruction methods also involve $B_0$ map estimation techniques leveraged to correct inhomogeneity artifacts. The corresponding first author and year of publication accompany each method. Lines joining different elements represent that they are related or that one technique has been incorporated into another.